\documentclass[twocolumn,prl]{revtex4}

\usepackage{graphicx}
\usepackage{bm}

\newcommand{\be}{\begin{equation}}
\newcommand{\ee}{\end{equation}}

\newcommand{\bfr}{ {\bf r} }

\begin{document}

\title{Ring exchanges and the supersolid phase of $^4$He}
\author{D. M. Ceperley}
\affiliation{Department of Physics and NCSA, University of
Illinois at Urbana-Champaign, Urbana, Illinois 61801}
\email{ceperley@uiuc.edu}
\author{B. Bernu}
\affiliation{Laboratoire de Physique Th\'eorique des Liquides, UMR
7600 of CNRS, Universit\'e P. et M. Curie, boite 121, 4 Place
Jussieu, 75252 Paris, France}

\begin{abstract}
Using Path Integral Monte Carlo we calculate exchange frequencies
in bulk hcp $^4$He as atoms undergo ring exchange. We fit the
frequencies to a lattice model and examine whether such atoms
could become a supersolid, that is have a non-classical rotational
inertia. We find that the scaling with respect to the number of
exchanging atoms is such that superfluid behavior will not be
observed in a perfect $^4$He crystal.
\end{abstract}

\maketitle

Recent torsional-oscillator observations by E. Kim and M. H. W.
Chan on solid $^4$He, both in the disordered absorbed
vycor\cite{chan1} and in bulk $^4$He \cite{chan2}, have revived
interest in the super-solid phase. In this phase, one has both
long range crystalline order and superfluidity. Experiments find
superfluid response to rotation at temperatures below 0.2K. The
low temperature superfluid density is about 1\% of the solid
density and is found to be independent of density within a
pressure range of 25 to 60Bars. In the following, we assume the
observed phenomena is described by equilibrium thermodynamics of
pure bulk helium. In practice, whether such is the case, is one of
the most important issues. Assuming equilibrium\cite{leggett}, a
supersolid is characterized by both long-range translational order
and by a non-classical response to rotation (NCRI).

In proposing such a state, Andreev and Lifshitz\cite{andreev}
postulated the existence of suitable defects ({\it e.g.}
vacancies) and then estimated the properties of the dilute system
of bosonic defects. The superfluid transition temperature of point
bosons is $T_c=3.31 \rho^{2/3}\hbar^2/m$ where $\rho$ is the
density of the bose condensing quasi-particles and $m$ their
effective mass. Taking the measured value for $T_c=0.2K$, writing
the mass in terms of the bare helium atom mass $m=\mu m_4$ and
their density as a fraction $c$ of the bulk helium density at
melting $\rho=0.029\AA^{-3}$ we find that $c\approx 0.012
\mu^{3/2}$: if the quasi-particle is heavier than a helium atom,
at least one out of every 83 lattice sites must have a defect; if
the mass is 1/10 the atom mass, the relative density of defects
will need to be $4\times 10^{-4}$. Such a dense collection of
defects could not be associated with the much lower concentration
of $^3$He impurities.

Experiments, particularly the NMR experiments\cite{guyer,meisel}
on solid $^3$He rule out zero point vacancy concentrations of more
than 1 part in 10$^{14}$.  Pederiva et al.\cite{pederiva} have
done calculations of the energy of vacancies and find an energy of
about 15 K$\pm$4K in the hcp phase of bulk $^4$He at the melting
density, and 30 K at 50 bars, in agreement with various
experiments, most notably X-Ray scattering\cite{simmons}. Using
path integral calculations (described below) we estimate the
energy to create an interstitial at the melting density is
48K$\pm$ 5K; they are even more improbable than vacancies. Hence,
the energies of point defects are substantially greater than zero;
in equilibrium there should not be enough of them at 0.2 K to bose
condense; this precludes the Andreev mechanism for a supersolid.

In this paper we examine the possibility that bulk solid hcp
helium, assumed to be free of defects such as impurities and
vacancies, could have a supersolid phase. One might think that
there would always be ground state defects, arising from the large
quantum zero point fluctuations. Near melting, the r.m.s.
vibration about the lattice site is 30\%, so that at any instant
of time, a good fraction of atoms are closer to a neighboring site
than to their home site. However, the absence of an atom from a
lattice site is not sufficient for having a supersolid;  if the
empty site is always accompanied by doubly occupied site, there
can no mass current. Chester\cite{chester} proved that any Jastrow
({\it i.e.} pair product) wavefunction of finite range has both
BEC and vacancies. However, Jastrow wavefunctions crystallize with
much difficulty and with a transition density off by an order of
magnitude\cite{hansen} so that they are not reliable enough to be
used to predict superfluidity. 
In general, variational wavefunction approaches are suspect, since
the energy is not sensitive to the low probability regions of
configuration space that are important for the superfluid density
and bose condensation.

Leggett\cite{leggett} has written an upper bound to the superfluid
density in terms of the zero temperature density of the solid.
Using densities from path integral calculations, we find that the
Leggett bound gives $\rho_s/\rho \leq 0.16$ at the melting
density, a value consistent with experiment. In this paper, we
calculate values for ring exchange frequencies and using them find
that the superfluid density should be zero in solid helium.

To determine whether bulk helium could be supersolid, path
integrals give a much cleaner frame-work; they can be used to
compute the superfluid density and the momentum distribution
without the assumption of a trial wave function or any other
uncontrolled approximation. The partition function of $N$ bosons
is:
 \be  Z= \frac{1}{N!} \sum_P \int dR \langle R | e^{-\beta H} |
 PR \rangle \ee
where $H$ is the Hamiltonian, $\beta$ the inverse temperature and
$R= \{ \bfr_1, \bfr_2 , \ldots \bfr_N \}$. For numerical
calculations, the density matrix operator is expanded into a path,
beginning at the configuration $R$ and ending at $PR$. In terms of
these paths, the superfluid density ({\it i.e.} the number of
atoms not moving with the walls of the torsional oscillator) is
given by:
 \be  \frac{\rho_s}{\rho} = \frac{m <\vec{W}\vec{W}>}{\hbar^2 \beta N } \ee
where $\vec{W}= \int_0^{\beta} dt \sum_{i=1}^N d\bfr_i(t)/dt $ is
the winding number of the path  around a torus. It is only
exchanges on the order of the sample size that contribute to the
superfluid density; local exchanges make no contribution. Using
the PIMC method, superfluidity and freezing, happen naturally at
the right density and temperature, without imposing them in any
way. The technical complications\cite{RMPI} concern ergodicity of
the random walk, and finite size effects: one has to take the
limit as $N \rightarrow \infty$.

PIMC calculations find a superfluid density on the order of 3\% at
melting density (molar volume 21.04 cm$^3$) and about 1.2\% at 55
bars (molar volume 19.01 cm$^3$) in a 48 atom ($3 \times 4 \times
2$) hcp supercell. The superfluid density is larger than what is
observed, and has a relatively weak pressure dependance. However,
a cell with 180 atoms, has zero superfluid density. We cannot be
sure that the lack of winding paths has a physical origin, or is
due to a lack of ergodicity within the PIMC random walk. To change
from one winding configuration to another involves a global move
of the paths which becomes very unlikely as the box size gets
large. To avoid this problem, we turn to a PIMC approach which
directly estimates individual exchange probabilities.

The Thouless\cite{thouless} theory of exchange in quantum crystals
assumes that at low temperatures, the system will almost always be
near one of the particular $N!$ arrangements of particles to
lattice sites with rare, rapid, tunnellings from one arrangement
to another.  
Hence, we label the particles with their initial lattice sites.
Then the partition function is written as a sum over permutations
of sites onto themselves.  We break up the permutation into cyclic
exchanges $\{ p_1,p_2 \ldots p_n\}$; each cycle is independent, so
that:
 \be \label{ITP} Z= Z_0  \sum_P\prod_{i=1}^{n_P} f_{p_i} (\beta) . \ee
where $Z_0=\int dR \langle R | e^{-\beta H} | R \rangle $ is an
uninteresting phonon partition function at low temperature. The
contribution for a cycle:
 \be f_p (\beta) = \frac{1}{Z_0}\int dR \langle R | e^{-\beta H} | p R \rangle \equiv J_p \beta.
  \ee
is proportional to $\beta$ because it is localized in time; the
coefficient, $J_p$ is the exchange frequency.


A special case of the above partition function is the
Feynman-Kikuchi (FK) model\cite{feynman} that assumes that $J_p=
J_0 e^{-\alpha L_p}$ where $L_p$ is the cycle length.  The
partition function is:
 \be  Z=  Z_0\sum_P  (\beta J_0)^{n_P} e^{ -\alpha \sum_{i=1}^{n_P} L_{p_i}}. \ee
We make contact with previous work by assuming $\beta J_0=1$.
Calculations were done assuming a cubic 3D lattice, first by
Feynman and Kikuchi analytically and then numerically by
Elser\cite{elser}. It is found that the superfluid transition
occurs as $\alpha$ becomes smaller from a localized state (small
cycles) to large cycles for $\alpha_c \approx 1.44$. To understand
this critical value, consider the free energy of adding a link to
an existing loop. A cubic lattice has coordination number 6 but a
return to the previously visited site is impossible since it is a
cyclic permutation. Adding a link costs probability $e^{-\alpha}$
but the entropy of the new link is $5$; hence $\alpha_c \sim
\ln(5)=1.6$. The critical value is smaller because of the
'self-avoiding' restriction within a cycle and on overlapping
cycles. Using the same argument for the hcp lattice gives a
critical coupling of $\alpha_c \approx 2.3$.

Using methods\cite{CJ,cornell} developed for solid $^3$He, we
calculate the exchange frequencies and estimate how close they are
to the critical value. We assume the helium interact with a
semi-empirical\cite{aziz95} pair potential. The frequencies for 2,
3 and 4 atom exchanges is very small\cite{RMPI}, {\it e. g.}
 $J_2\sim 3 \mu K $ at melting density.
However, small cyclic exchanges are quite different from the long
exchanges needed to get a supersolid. We have performed exchange
calculations of 50 different exchanges involving from 5 to 10
atoms. All exchanges involve nearest neighbors, since calculations
show that next-nearest exchanges are much less probable. We obtain
accuracies on the order of 5\% for the 5-particle exchanges and
10\% for the 9 particle exchanges. More than half of the exchanges
involve winding around the cell boundaries, important because they
are representative of the type of exchanges in a supersolid.

\begin{figure}
\includegraphics[width=8cm,height=8cm]{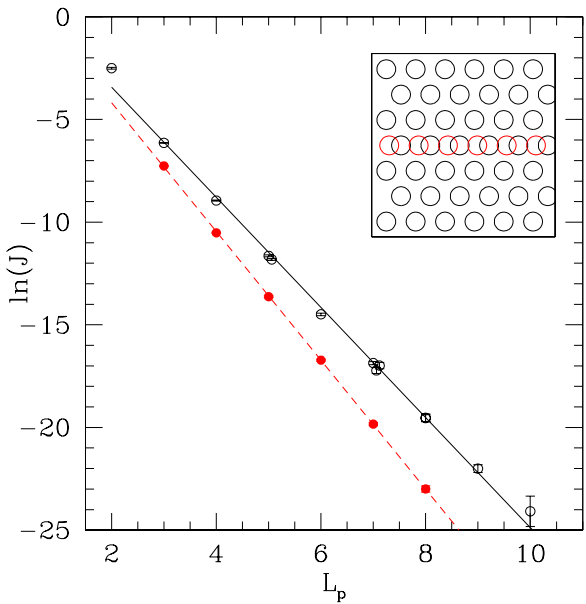}
\caption{\label{winding} The exchange frequencies (J in K) versus
exchange length $L_p$ for straight line exchanges in the basal
plane that wind around the periodic cell. The inset shows the
lattice sites in the basal plane for $L_p=6$; the red atoms show
the atoms midway through the exchange. The circles with error bars
are the $\ln(J)$ at two molar volumes; 21.04 cm$^3$ (black, open
circles, solid line) and 19.01 cm$^3$(red, solid circles, dashed
line). Multiple points at the same value of $L_p$ are from cells
with different number of atoms in the directions perpendicular to
the exchange direction. The lines are least squares fits.}
\end{figure}

Fig. \ref{winding} shows the results of calculations of the
frequency of the simplest winding exchanges: straight line
exchanges in the basal plane. As assumed in the FK model, we find
that the exchange frequencies decrease exponentially with the
length of the exchange with an exponent of  $\alpha =2.64$ near
the melting volume 21.04 cm$^3$, and $\alpha=3.14$ at the volume
19.01 cm$^3$ corresponding to $P \sim$ 60 bars.

\begin{figure}
\includegraphics[width=8cm,height=8cm]{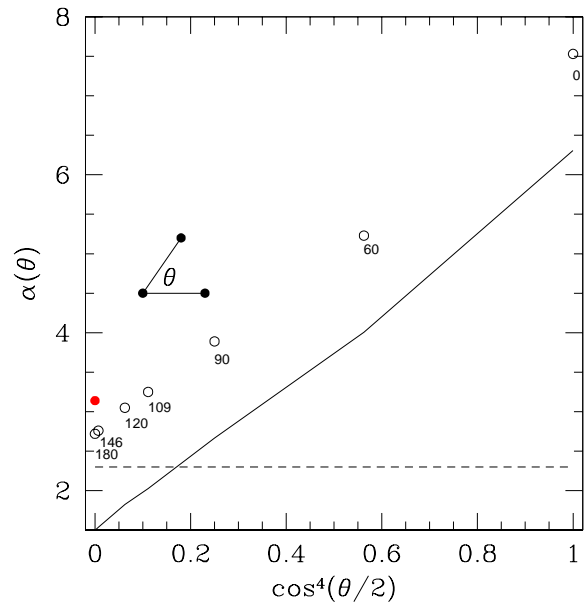}
\caption{\label{2pfit} (Color online) The model parameters vs.
$\cos^4(\theta/2)$ at the melting density (open circles). The
angle is the interior angle of the vertex as shown. The fits were
done using 50 exchanges in a 180 atom cell. The solid red circle
corresponds to a pressure 60 bars. The two lines show a critical
supersolid Hamiltonian with (solid line) and without (dashed line)
angular dependance. }
\end{figure}

To construct a more realistic model than the FK model, we need to
take into account more details of the geometry of the exchange
than just the number of exchanging atoms. We assume that it is the
internal geometry of the exchange that matters\cite{lesh}; the
detailed arrangement of the neighboring spectator atoms is much
less important. In particular, we assume the action of a given
n-cycle (the log of the exchange frequency) is the sum over the
internal vertices of the exchange:
 \be \label{model} J_p=J_0  \exp [-\sum_{k=1}^p\alpha(\theta_k)].\ee
Here $\theta_k$ are angles between successive displacements in the
exchange cycle. In an hcp lattice there are 7 possible angles
between two nearest neighbor displacement vectors, but $\theta=0$
only occurs in the pair exchange. 
We determine the parameters by fitting to the PIMC exchange
frequencies $J_p$. Good fits are obtained; the model predicts the
exchanges frequencies with an accuracy of about 20\%. The
resulting coefficients are shown in Fig. \ref{2pfit}. Typical
errors on the coefficients are $\pm 0.03$. Note that there is a
strong preference for exchanges that proceed in a straight line
versus ones which have sharp angles; for them the incoming and
outgoing particles are more likely to collide. The fitting
coefficients are seen to depend linearly on $\cos^4(\theta/2)$. We
find no preference for exchange in the basal plane beyond the
effect induced by the selection of angles.

We also verified this model by comparing with results of the
direct PIMC calculations (where permutations are generated
dynamically) mentioned above and obtained rough agreement with the
winding numbers and permutation cycle distributions. The 2- and 3-
particle exchanges are about twice the prediction of model. On the
other hand, the exchanges in the 180 atom cell are somewhat
smaller than Thouless theory. We expect such disagreements because
Eq. \ref{model} does not contain the effect of spectator atoms on
the exchange process, and small exchanges have not been included
in the fit.

The activation coefficient of the exchange is $J_0=7.2 \pm 0.5K $.
This is the order of magnitude but smaller than the vacancy
energy. If it is related to vacancies, then $J_0$ should be of the
same order of the vacancy energy at all densities, but on the
contrary, we do not find a significant density dependance.

We now analyze the model. Because the probability of retracing is
small, we initially neglect the self intersections. This leads to
a diffusion problem on a lattice, with a large probability of
continuing in the same direction.
Since a displacement only depends on the previous displacement, it
is an un-normalized Markov process. (Because the hcp lattice has a
basis, we have to label the displacements consistently on the A
and B planes so that the transition probabilities are independent
of the plane.) The probability of a given displacement vector
$\Pi_i$  will approach a steady state after many steps, ($i$
refers to one of the lattice directions) and satisfy the
eigenvalue equation:
 \be \sum_{i} \Pi_i e^{\alpha_{ij}}=\lambda \Pi_j. \ee
By symmetry $\Pi_i$ can only depend on whether the direction is in
the basal plane, $\Pi_1$, or out of the basal plane, $\Pi_2$.
Define the partial sum $D_{i,j}=\sum_{j}e^{-\alpha_{ij}}$ where
$(i,j)$ are either in the basal plane (1) or out of it (2), {\it
e.g.} $D_{11}$ is the sum of the probability of both vectors being
in the basal plane). The eigenvalue $\lambda$, is the solution to
the secular equation:
 \be ( D_{11}-\lambda)(D_{22}-\lambda)-D_{12}^2=0. \ee
Putting in the model parameters, we find $\lambda=
 0.303 \pm 0.005$.
The probability of having an exchange of length $p$ will equal a
prefactor times $\lambda^p$, hence,  since $\lambda<1$, our PIMC
results imply that solid $^4$He  will have only localized exchange
and thus cannot be a supersolid.


Let us consider how the neglect of the self-intersections affects
the critical value. Qualitatively, self-intersections must
decrease the probability of long exchange cycles. To estimate the
effect quantitatively, we perform random walks on the hcp lattice
and only count non-intersecting walks. 
If we artificially change the model parameters $\alpha$ by
subtracting $1.2$, the model becomes critical when $\lambda
\approx 1.03$ as opposed to $\lambda=1$. This line is shown in
figure 2. Our computed value of $\lambda$ is much less than what
is needed to allow for a supersolid. We expect that the presence
of other exchange cycles will further increase the critical value
of $\lambda$.


In summary, PIMC-computed exchange frequencies for hcp solid
$^4$He show that only localized exchanges will be present and thus
should not exhibit the property of nonclassical rotational
inertia. Based on other theoretical and experimental findings, we
think it unlikely that the observed phenomena are due to
vacancies, interstitials or $^3$He impurities. Hence, one must
look for an explanation of the experiments elsewhere, either to
non-equilibrium effects or more complicated lattice defects.

This research was supported by the NASA program in Fundamental
Physics and the Department of Physics at the University of
Illinois Urbana-Champaign.

\end{document}